\documentclass[pra,preprint,showpacs,showkeys]{revtex4-1}
\usepackage{graphicx}
\usepackage[latin1]{inputenc}

\begin{document}
\title{Spectroscopy of PTCDA attached to rare gas samples: clusters vs. bulk matrices. II.~Fluorescence emission spectroscopy}

\author{Matthieu Dvorak$^1$, Markus M\"uller$^1$, Tobias Knoblauch$^2$, Oliver B\"unermann$^{1,3}$, Alexandre Rydlo$^4$, Stefan Minniberger$^4$, Wolfgang Harbich$^4$, and Frank Stienkemeier$^1$}
\affiliation{
$^1$ Physikalisches Institut, Universit\"at Freiburg, Hermann-Herder-Str.~3, D-79104 Freiburg, Germany\\
$^2$ 1. Physikalisches Institut, Universit\"at Stuttgart, Pfaffenwaldring 57, 70550 Stuttgart, Germany\\
$^3$ Institut f\"ur Physikalische Chemie, Georg-August-Universit\"at, Tammannstr. 6, 37077 G\"ottingen, Germany\\
$^4$ Institut de Physique des Nanostructures, École Polytechnique Fédérale de Lausanne (EPFL), CH-1015 Lausanne, Switzerland}



\begin{abstract}
The interaction between PTCDA (3,4,9,10-perylene-tetracarboxylic-dianhydride) molecules and solid rare gas samples is studied by means of fluorescence emission spectroscopy. On the one hand, laser-excited PTCDA-doped large argon, neon and para-hydrogen clusters in comparison with PTCDA embedded in helium nanodroplets are spectroscopically characterized with respect to line broadening and shifting. A fast non-radiative relaxation is observed before a radiative decay in the electronic ground state takes place. On the other hand, fluorescence emission studies of PTCDA embedded in bulk neon and argon matrices results in much more complex spectral signatures characterized by a splitting of the different emission lines. These can be assigned to the appearance of site isomers of the surrounding matrix lattice structure.
\end{abstract}

\date{\today}
\maketitle

\section{Introduction}
The spectroscopy of molecules is most of the time an environment-dependent measurement since many species cannot be probed under gas phase conditions. The broadening of the spectral features due to interaction with the substrate on which the molecule is deposited, or with a solvent or matrix in which the molecule is embedded is unavoidable. Furthermore, broadening effects induced by the environment temperature are even more complicating the measurements and their interpretation.

One way to limit both of these effects is to use rare gas matrices. The interaction between a chromophore and, e.g., an argon or neon matrix is much weaker than for other solid substrates and their achievable low temperatures contribute to a drastic improvement in spectroscopic resolution. For this reason many studies on rare gas matrices doped with atoms and smaller molecules have already been published (a compendium of these molecules and their spectral features can be found in Ref.~\cite{Jacox2003}). Some larger polycyclic aromatic hydrocarbon (PAH) molecules and their ions, embedded in argon or in neon matrices have already been studied: C$_{60}$~\cite{Sassara1996}, naphthalene~\cite{Salama1991}, phenanthrene~\cite{salama1994}, pentacene~\cite{Halasinski2000}, 9,10-dichloroanthracene~\cite{crepin1990}, benzo[g,h,i]perylene~\cite{Chillier2001} and perylene~\cite{Joblin1995,Joblin1999}. However, in spite of the improvements resulting from the use of rare gas matrices, the effect of the matrix environment 
which stabilizes the guest molecule, is still prominent when compared to gas phase studies. This affects mostly the electronic but also the vibrational spectra. The matrix effects on the outcome of a spectrum can be divided in three categories: inhomogeneous broadening, shifting and splitting of lines.\cite{Almond1991,Bondybey1996,Khriachtchev2011}

As a different approach the interaction can be weakened by supporting the guest molecule on a rare gas surface. This challenging technique can alternatively be realized by using large rare gas clusters instead of bulk matrices. Indeed, due to their solid state, the doping of such clusters after their formation via a pickup process limits the possible localization of the chromophore to surfaces sites.\cite{Dvorak_P1} Restricting the occupation of the host cluster to one guest molecule provides samples with isolated molecules. Multiple doping of rare gas clusters in order to increase target density carries the risk to form dimers or oligomers. Indeed, atoms and small molecules ($<10$ atoms) are mobile on the surface of rare gas clusters~\cite{Lallement1992} so that in the case of heterogeneous doping of clusters, it could be shown that these clusters can serve as a chemical reaction center for the formation of complexes.~\cite{Briant2000} More recently the formation of barium-xenon exciplexes on the surface of 
large argon clusters could be demonstrated.~\cite{Briant2010} However recent measurements in our group pointed out that this surface mobility does not apply to larger PAHs such as PTCDA and that the molecules stay isolated from each other.~\cite{Dvorak_P1}

The chromophore used in the present work is the PTCDA molecule (3,4,9,10-perylene-tetracarboxylic-dianhydride, C$_{24}$H$_{8}$O$_{6}$). The ability of this perylene derivative to form highly ordered structures on surfaces~\cite{Schreiber2000,Qing2009} made it very popular in opto-organic applications and resulted in a large number of spectroscopic studies on a variety of substrates. Moreover, molecules of this type embedded in helium nanodroplets
have already been extensively characterized in our group.~\cite{Wewer2003, Wewer2004, Wewer2005, Roden2010}

The present article is the second part of the spectroscopic study of PTCDA attached to rare gas samples and is devoted to fluorescence emission measurements (absorption and excitation measurements are presented in Ref.\,\cite{Dvorak_P1}, later referred to as Paper I). It is structured as follows: first, the fluorescence emission spectroscopy of single PTCDA molecules attached to large argon, neon or para-hydrogen clusters is presented. These complexes are excited at different vibrational levels in order to give an insight on the relaxation process of such complexes. These measurements are compared to those obtained with PTCDA-doped superfluid helium droplets. Then, the fluorescence emission spectroscopy of PTCDA embedded in argon and neon matrices is presented and compared with absorption measurements. The comparison between rare gas clusters and rare gas matrices allows to comment on effects by the crystal structure and of surface vs.~inside localization of the chromophore on the spectral signatures. The 
presented measurements also bridge the gap between measurements obtained in the gas phase or with rare gas clusters and those obtained on surfaces or in the bulk.

\section{Experimental Setup}\label{setup}
The experimental setup used for the production of clusters as well as their doping with PTCDA and their spectroscopic study has been thoroughly presented in Paper I and is here only briefly described. Large argon, neon and para-hydrogen clusters as well as helium droplets are obtained via supersonic expansion of high-pressurized gas through a cold nozzle (a few tens of $\mu$m). Normal hydrogen is converted into para-hydrogen with the help of an in situ catalytic converter. However, in the course of the measurements presented here no differences in the spectral features have been found dependent on the different spin entities of the hydrogen molecule. With the expansion parameters used, the argon, neon and para-hydrogen clusters have sizes ranging from 10$^2$ to 10$^4$ atoms; all of the formed clusters are solid in contrast to helium droplets which are superfluid (mean size about 20\,000 atoms per droplet).\cite{Grebenev1998} The clusters are doped via a pickup process when traversing a heated cell containing 
an adjustable partial pressure of PTCDA (Sigma Aldrich, used without further purification).

The doped clusters are excited by tunable laser radiation (resolution $<$\,0.1\,cm$^{-1}$) and the resulting fluorescence emission light is imaged onto an optical fiber bundle. The exit of this fiber bundle is mounted at the entrance slit of a spectrograph (Jobin-Yvon: SPEX 270M, grating: 1200 lines/mm). A telescope adapts the numerical aperture of the bundle with the one of the spectrograph. The emitted signal is recorded by a charge coupled device camera (CCD, Andor, Newton) fixed on the exit aperture of the spectrograph. The resolution of the setup was measured to be 18\,cm$^{-1}$~Full-Width at Half Maximum (FWHM).

Rare gas bulk matrices are produced by co-deposition of rare gas atoms and PTCDA molecules on a cold substrate (8\,K for neon and 28\,K for argon matrices; the other deposition parameter are discussed in Paper I). The matrices have a typical thickness of about 50\,$\mu$m and a surface of a few square mm. Fluorescence measurements are obtained by focusing a laser beam (wavelength 473\,nm) perpendicularly to the matrix surface. The fluorescence light is collected by an optical fiber placed at the side of the matrix perpendicularly to the laser beam. The output of the fiber is coupled to a spectrograph (Jobin-Yvon T64000, CCD Spectrum One). This arrangement ensures that only the fluorescence light is collected. The spectral resolution of the system using a 2400 lines/mm grating is 0.5\,cm$^{-1}$.

\section{Experimental results}\label{EM_results}
\subsection{Emission spectra of PTCDA-doped clusters}
\subsubsection{Emission of PTCDA embedded in helium nanodroplets}
\begin{figure}
{\center\includegraphics[width=8.5cm]{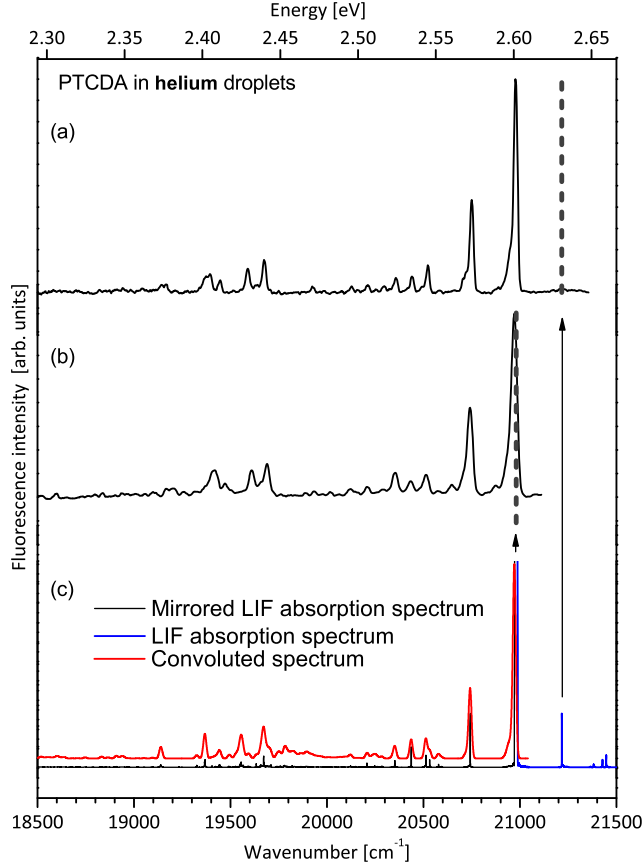}}
  \caption{Fluorescence emission spectra of PTCDA molecules embedded in helium droplets. Two excitation energies were probed: the 0-0 line of the $S_1 \leftarrow S_0$ transition at 20\,987.8\,cm$^{-1}$~(b) and the first vibration at 21\,217.3\,cm$^{-1}$~(a). These spectra are identical to the mirrored LIF spectrum (c), convoluted with a Gaussian distribution (FWHM = 18\,cm$^{-1}$) taking into account the inferior resolution of the spectrograph-CCD system. }
  \label{EM_helium}
\end{figure}
Fig.~\ref{EM_helium} shows the fluorescence emission spectra of single PTCDA molecules embedded in helium droplets. The spectra are recorded at two different excitation energies, at 20\,987.8\,cm$^{-1}$~(Fig.~\ref{EM_helium}~(b)) corresponding to the 0-0 line of the $S_1 \leftarrow S_0$ transition and at 21\,217.3\,cm$^{-1}$~(Fig.~\ref{EM_helium}~(a)) corresponding to the first vibrational mode. Spectra are compared to the Laser-Induced Fluorescence (LIF) absorption spectrum mirrored with respect to the 0-0 line and artificially broadened with a Gaussian distribution taking into account the inferior resolution of the optical system. The fluorescence emission spectra are to a large extent identical to the mirrored absorption spectrum. The relative positions of the different vibronic lines only show weak deviations for modes at lower energies, which are due to discrepancies in the calibration procedure. The similarity between the fluorescence emission spectra and the mirrored absorption LIF spectrum means that 
electronic ground and excited potential energy curves are quite the same. This was already clear from the comparison between the LIF and Raman spectra of PTCDA.\cite{Wewer2004,scholz_raman} The relative intensities are also very similar in both fluorescence emission spectra and in the mirrored LIF spectrum except for the 0-0 transition which has slightly more intensity in the absorption spectrum. The line widths of the emission spectra are determined by the spectrometer, so only an upper limit can be deduced from the spectra. Previous studies on e.g.~Mg-phthalocyanine embedded in helium droplets showed a typical FWHM of about 1\,cm$^{-1}$, comparable to the LIF absorption measurements.\cite{lehnig2003} One might expect a similar behavior for PTCDA molecules probed in the present experiment.

The excitation of the first vibronic mode (Fig.~\ref{EM_helium}~a) does not result in a fluorescence at this energy (dashed line in Fig.~\ref{EM_helium}~(a)) but leads to an unshifted spectrum when compared to the 0-0 excitation. This indicates a fast non-radiative process into the ground level of the excited state $S_1$ from which the fluorescence process into the ground state $S_0$ takes place. We measured the radiative lifetime for this transition to 5.5$\pm$0.5\,ns (see Ref.\,\cite{Buenermann2006} for the method).

From the results on fluorescence emission spectroscopy of Mg-phthalocyanine a similar process has been concluded. However, for that molecule the excitation of higher vibrational modes results in a splitting of the lines in the fluorescence spectrum. This was interpreted not to be originated from radiating higher vibronic levels, but related to different helium solvation configurations around the molecule.\cite{lehnig2003} Such a splitting, if present, would not be observable in our experiment because of the spectral resolution of the instrument.

\subsubsection{Large para-hydrogen clusters}
\begin{figure}
{\center\includegraphics[width=8.5cm]{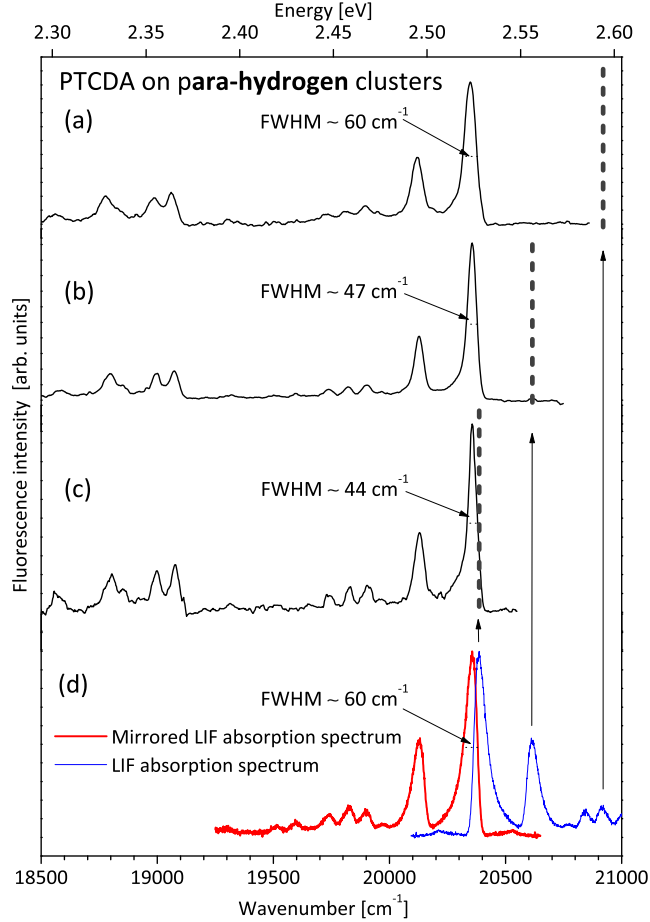}}
  \caption{Fluorescence emission spectra of PTCDA molecules attached to large para-hydrogen clusters. Three excitation energies (dashed lines) are probed: a radial stretching mode of C-C bonds at 20\,920\,cm$^{-1}$~(a), the first vibration at 20\,615\,cm$^{-1}$~(b) (breathing mode) and the 0-0 line at 20\,386\,cm$^{-1}$~(c).}
  \label{EM_hydro}
\end{figure}
Similar to the measurements on helium droplets, fluorescence emission spectroscopy of PTCDA attached to large para-hydrogen clusters has been performed. In Fig.~\ref{EM_hydro} the results are displayed for three different excitation energies. The emission spectrum obtained upon the excitation of the 0-0 line at 20\,386\,cm$^{-1}$~(Fig.~\ref{EM_hydro}\,(c)) is identical to the mirrored LIF spectrum (d). Both the relative positions and relative intensities match. The only difference observed is the spectral resolution. A FWHM of about 44\,cm$^{-1}$, identical for each peak is obtained, which is notably smaller than the value obtained in the LIF absorption spectrum (60\,cm$^{-1}$), but still quite larger when compared to the resolution of the instrument. An extra broadening mechanism in the absorption spectrum is expected to originate from the strongly reduced lifetime due to the vibrational relaxation which increases at higher excitation energies due to the increasing number of relaxation channels.

The fluorescence emission spectrum for the excitation of the first vibration of the $S_1$ state is shown in Fig.~\ref{EM_hydro}\,(b). The excess energy of 230\,cm$^{-1}$~compared to the 0-0 transition is, as in the helium spectrum, completely dissipated prior to the radiative decay. The only difference observed is a slight increase of the FWHM to 47\,cm$^{-1}$. For an even larger excess energy value (534\,cm$^{-1}$), which corresponds to vibrational mode at 20\,920\,cm$^{-1}$~in the LIF spectrum (radial stretching mode of C-C bonds; Fig.~\ref{EM_hydro}\,a), the spectrum is again identical to the others, now having an increased FWHM of about 60\,cm$^{-1}$. Taking into account the broadening coming from the spectrograph, still the effective line widths are smaller compared to the absorption measurements. An increase of the FWHM with the excess energy was also observed in the fluorescence emission spectra of perylene in an argon supersonic jet.\cite{Imasaka1984} In that case, it was observed that an excess 
energy of 1\,600\,cm$^{-1}$~results in an increase of the FWHM of the peaks from $\sim$\,10 to $\sim$\,220\,cm$^{-1}$. This broadening of the optical features with the excess excitation energy was interpreted by an increased number of relaxation paths to the ground state and/or due to extra non-radiative decay channels in the electronically excited state.

Similar to the measurements in helium droplets, no hot bands are observed upon the excitation of higher vibronic levels. Therefore, also the vibrational excited molecules attached to para-hydrogen clusters exhibit a fast non-radiative decay into the vibrational ground level of the $S_1$ state, followed by a radiative transition into the $S_0$ state. It has to be noted that we see no indication for any fluorescence quenching by the hydrogen environment, as has been observed in alkalis attached to hydrogen clusters.\cite{Callegari1998}

\subsubsection{Large argon and neon clusters}
\begin{figure}
{\center\includegraphics[width=8.5cm]{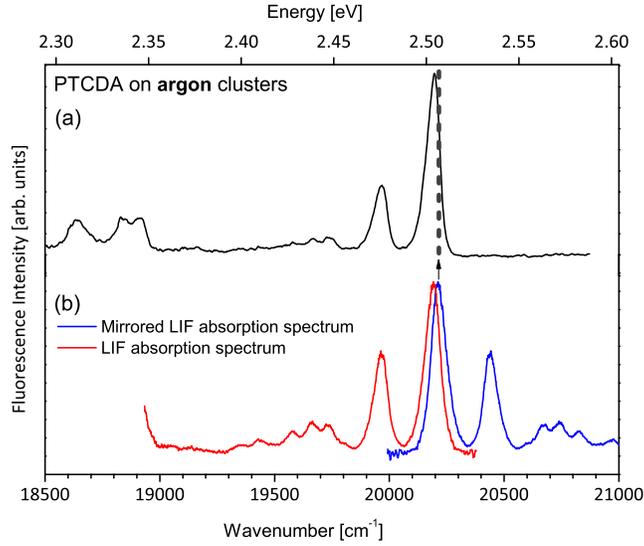}}
  \caption{Fluorescence emission spectra of PTCDA molecules attached to large argon clusters (a) compared with the mirrored LIF absorption spectrum (b). Excitation at the 0-0 line at 20\,216\,cm$^{-1}$~is probed.}
  \label{EM_argon}
\end{figure}
\begin{figure}
{\center\includegraphics[width=8.5cm]{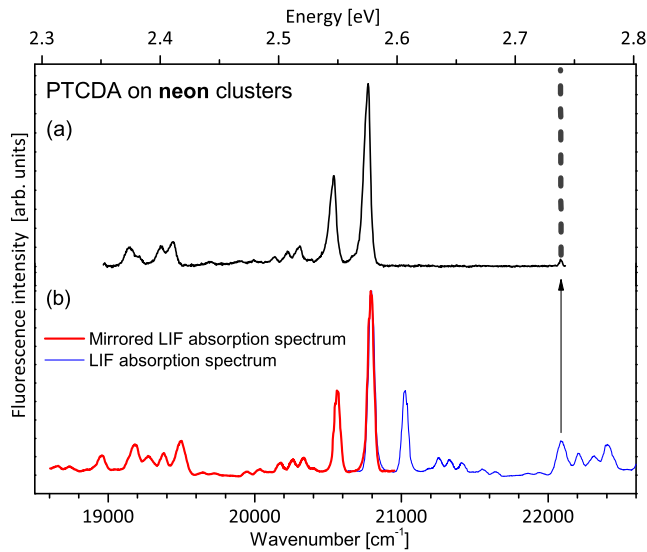}}
  \caption{Fluorescence emission spectra of PTCDA molecules attached to large neon clusters for the excitation energies corresponding to the vibronic mode at 22\,088\,cm$^{-1}$~(a) in comparison with the mirrored LIF absorption spectrum (b). The small peak at 22\,088cm$^{-1}$~can be assigned unambiguously to stray light from the laser beam.}
  \label{EM_neon}
\end{figure}

The fluorescence emission spectrum of PTCDA attached to large argon clusters is displayed in Fig.~\ref{EM_argon}\,(a) for the excitation of the 0-0 transition (20\,216\,cm$^{-1}$). This spectrum is compared to the mirrored LIF absorption spectrum (Fig.~\ref{EM_argon}(b)). For PTCDA attached to large neon clusters, four fluorescence emission spectra have been recorded, for four different excitation energies: the 0-0 transition at 20\,797\,cm$^{-1}$, the first vibration at 21\,024\,cm$^{-1}$, a radial stretching mode of C-C bonds at 21\,330\,cm$^{-1}$~and a deformation mode of C-H bonds at 22\,088\,cm$^{-1}$. The four recorded spectra are identical so only the one corresponding to the vibronic mode at 22\,088\,cm$^{-1}$~is displayed in Fig.~\ref{EM_neon}~(a). For both argon and neon clusters the relative positions and relative intensities of the different peaks are independent of the excitation energy. In contrast to the measurements with para-hydrogen, only a slight increase of the FWHM of a few wavenumbers 
with the excess energy is observed. Note that for argon as for neon the line width remains below those of the corresponding absorption measurements. Furthermore, neon clusters exhibit significant less broadening when compared to argon which is expected from the weaker interaction. Again, the excitation of higher vibronic levels does not result in any hot band emission. The small peak at 22\,088cm$^{-1}$~in Fig.~\ref{EM_neon}~(a) is unambiguously assigned to the Rayleigh scattering of the excitation laser on the cluster beam and is also observed with an identical intensity for an undoped neon cluster beam.

\subsection{Fluorescence in a rare gas bulk matrix}
\subsubsection{Neon matrix}
\begin{figure}
{\center\includegraphics[width=8.5cm]{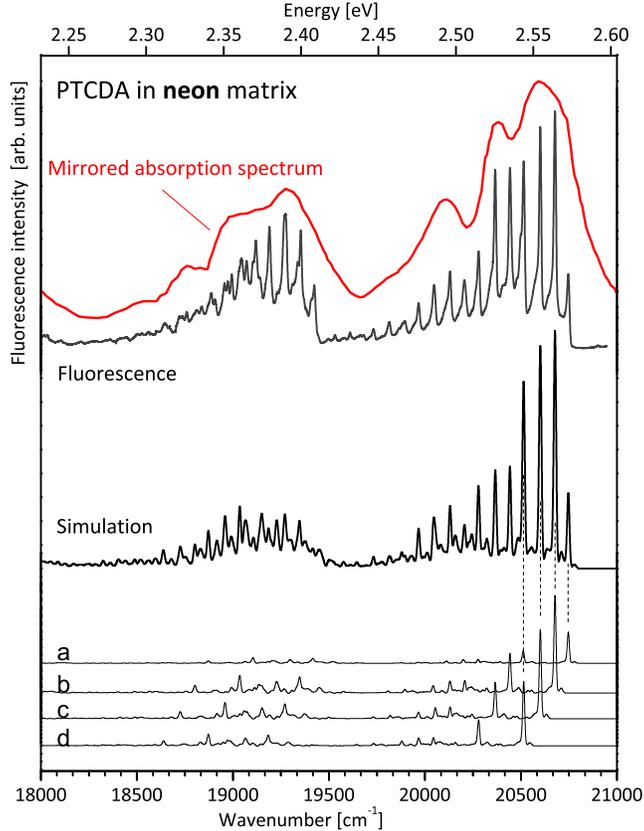}}
  \caption{Fluorescence emission spectrum of PTCDA molecules embedded in a neon matrix (top) compared with the mirrored absorption spectrum of the same matrix (red line). This spectrum can be reproduced (middle) by mixing four replicas of the excitation spectrum of PTCDA in the helium droplet with well-adjusted shifts and weighting factors (bottom).}
  \label{Fluo_neon}
\end{figure}
The fluorescence emission spectrum of PTCDA embedded in a neon matrix is displayed in the upper part of Fig.\,\ref{Fluo_neon} and is much more complex when compared to the just introduced spectra of molecules attached to clusters. The absorption spectrum of the same sample is also shown (red line), mirrored with respect to the laser excitation line at 21\,140\,cm$^{-1}$~and shifted to the red by 101\,cm$^{-1}$~for easier comparison. The first striking observation is the increase of the spectral resolution by more than one order of magnitude for the emission spectrum (FWHM of about 16\,cm$^{-1}$) when compared to the absorption spectrum. The broadening of the latter is thought to be due to overlapping of individual vibrational bands in the excited states of different site isomers and extra line broadening as discussed above.

Apart from a different broadening, both the mirrored absorption spectrum and the fluorescence spectrum show the same profile. Each subfeature of the mirrored absorption spectrum can be related to different peaks of the fluorescence spectrum. The interaction of the molecule with the neon host leads to splitting, shifting and broadening of lines. The PTCDA molecule apparently can bind to a well defined number of different sites in the neon host resulting in different
shifts.\cite{Joblin1999} Unlike small embedded atoms or molecules, a large molecule substitutes a considerable number of host atoms leading to a significant number of host guest combinations. Such site isomers are characterized by different interaction strength with the matrix, resulting in a splitting of the spectral response into multiple components having different relative intensities depending on their relative abundance. The absorption spectra of perylene, a similar molecule to PTCDA, has been measured in rare gas matrices too and showed also a splitting of the main features. Each main
transition was split in two components in a neon matrix, in five in an argon matrix and six in a krypton matrix, respectively.\cite{Joblin1999,Biktchantaev2002}.

Following the idea of splitting, the spectrum can be reproduced by superposing identical replicas of a reference spectrum of PTCDA (Fig.~\ref{Fluo_neon}, middle). The reference spectrum used is the LIF spectrum of PTCDA embedded in helium droplets (see Fig.\ref{EM_helium}\,c) which has been mirrored with respect to the 0-0 line of the $S_1 \leftarrow S_0$ transition. The mirrored spectrum has been convoluted with a Gaussian distribution to allow for matrix broadening. In solid neon, only four replicas, shifted and with different intensities, are sufficient to reproduce the experimental spectrum. In this way the spectrum can be well simulated. The four replicas --- a to d at the bottom of Fig.~\ref{Fluo_neon}~--- used to reproduce the neon matrix absorption spectrum are red-shifted by 241, 310.5, 387 and
473.5\,cm$^{-1}$~ with respect to the 0-0 line of PTCDA in helium droplets at 20\,987.8\,cm$^{-1}$, respectively, and have weighting factors of 11\,\%, 34\,\%, 32\,\% and 23\,\%, respectively. The typical error in the matrix spectrum is of $\pm$~0.5\,cm$^{-1}$. Note that the less intense peaks, in particular in the range between 20\,300 and 20\,000\,cm$^{-1}$~would ask for an increased number of replicas. Furthermore, some discrepancies in the relative intensity of some peaks (between 19\,200 and 19\,300\,cm$^{-1}$~and around 20\,400\,cm$^{-1}$) cannot be understood in our simplified simulation. Still, the overall agreement
is satisfactory for an unambiguous assignment of each of the main features observed to different vibration modes of different site isomers of PTCDA molecules embedded in the neon matrix.

\subsubsection{Argon matrix}
\begin{figure}
{\center\includegraphics[width=8.5cm]{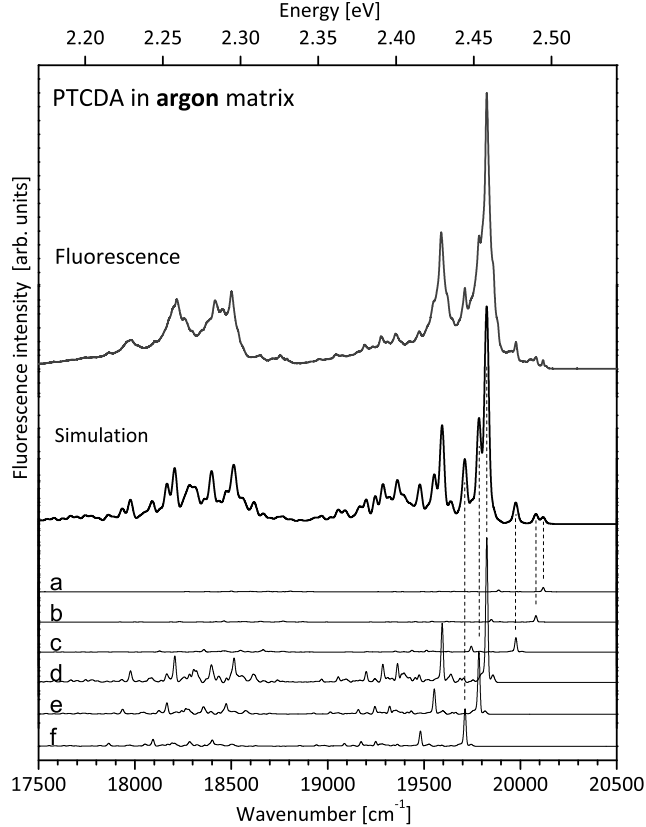}}
  \caption{Fluorescence emission spectra of PTCDA molecules embedded in an argon matrix (top). This spectrum can be reproduced (middle) by mixing six replicas of excitation spectrum of PTCDA in the helium droplet with corresponding shifts and weighting factors (bottom).}
  \label{Fluo_argon}
\end{figure}
The fluorescence emission spectrum in a solid argon matrix is shown in Fig.~\ref{Fluo_argon} and evidently demonstrates the stronger host-guest interaction by an enhanced red-shift of roughly 800\,cm$^{-1}$~when compared to the neon matrix spectrum. The relative shift is quite similar to that found for perylene (788\,cm$^{-1}$).\cite{Joblin1999} The broadening of the individual constituents is about 25\,cm$^{-1}$~which also is significantly enhanced when compared to neon (16\,cm$^{-1}$). The larger perturbation of the spectra can be related to a stronger interaction between the chromophore and the matrix,
mainly caused by a larger polarizability.\cite{Biktchantaev2002,Jacox2003}

Beside the interaction strength, the matrix temperature is known to play an important role on line widths. Crépin \emph{et al.} showed for 9,10-dichloroanthracene in argon matrices that the typical line width varies between 12\,cm$^{-1}$~at 12\,K and up to 30-40\,cm$^{-1}$~at 25\,K.\cite{crepin1990} Due to the widths of the different peaks and their proximity to each other it is not possible to evaluate the line shapes and their possible asymmetry. In the present studies we were not able to record an absorption spectrum due to the weak transmission of our argon matrix caused by smaller grain sizes compared to neon matrices resulting in stronger light scattering.\cite{Dvorak_P1,Klein1976}

Employing the same procedure as for neon, the argon spectrum can be reproduced with six replicas of the mirrored LIF spectrum of PTCDA in helium droplets convoluted with a Gaussian distribution  25\,cm$^{-1}$~in width. These (cf.~a to f at the bottom of Fig.~\ref{Fluo_argon}) have red-shift values of 869, 907, 1011, 1162, 1203 and 1275\,cm$^{-1}$, again compared to the origin of the $S_1 \leftarrow S_0$ transition of PTCDA in helium droplets at 20\,987.8\,cm$^{-1}$. Weighting factors are 1.6\,\%, 2.4\,\%, 5.4\,\%, 53.6\,\%, 23.1\,\% and 13.9\,\%, respectively. It should be noted here that most of
the intensity is due to only 3 replicas, giving more than 90\,\% of the amplitude of the simulated spectrum. One might speculate that due to the higher temperature of the argon matrix (28\,K, neon: 8\,K), rearrangement effects are more favorable, resulting in a higher population of low energy conformations.\cite{Joblin1999}

\section{Conclusion}
This paper presents fluorescence emission spectra of PTCDA attached to weakly interacting media. Single PTCDA molecules were attached to large argon, neon and para-hydrogen clusters. The excitation of the 0-0 line of the $S_1 \rightarrow S_0$ transition results in similar fluorescence spectra for the different clusters used, only distinguished by different red-shifts and line widths which are characterized in this study. Comparison with the mirrored excitation spectrum (LIF) shows an identical vibrational structure. The excitation of higher vibrational modes does not produce new spectral
features, pointing to a fast relaxation of vibrational states compared to radiative processes. Furthermore fluorescence emission spectra of PTCDA molecules embedded in argon and neon matrices have been recorded. The resolution of such spectra proved to be one order of magnitude better than corresponding absorption spectra. The many observed fluorescence lines have been successfully assigned to vibrational modes of different site isomers of PTCDA in the rare gas matrices. In conclusion, the combination of cluster and matrix spectra together with the quite well resolved emission spectra of doped matrices provide a conclusive assignment of the measured intensities. In particular also the quite broad absorption
spectra can be understood within the framework of the inhomogeneity of different site isomers.

\section{Acknowledgements}
Fruitful discussions with Takamasa Momose and Andrey Vilesov are gratefully acknowledged.


\begin{thebibliography}{30}%
\makeatletter
\providecommand \@ifxundefined [1]{%
 \@ifx{#1\undefined}
}%
\providecommand \@ifnum [1]{%
 \ifnum #1\expandafter \@firstoftwo
 \else \expandafter \@secondoftwo
 \fi
}%
\providecommand \@ifx [1]{%
 \ifx #1\expandafter \@firstoftwo
 \else \expandafter \@secondoftwo
 \fi
}%
\providecommand \natexlab [1]{#1}%
\providecommand \enquote  [1]{``#1''}%
\providecommand \bibnamefont  [1]{#1}%
\providecommand \bibfnamefont [1]{#1}%
\providecommand \citenamefont [1]{#1}%
\providecommand \href@noop [0]{\@secondoftwo}%
\providecommand \href [0]{\begingroup \@sanitize@url \@href}%
\providecommand \@href[1]{\@@startlink{#1}\@@href}%
\providecommand \@@href[1]{\endgroup#1\@@endlink}%
\providecommand \@sanitize@url [0]{\catcode `\\12\catcode `\$12\catcode
  `\&12\catcode `\#12\catcode `\^12\catcode `\_12\catcode `\%12\relax}%
\providecommand \@@startlink[1]{}%
\providecommand \@@endlink[0]{}%
\providecommand \url  [0]{\begingroup\@sanitize@url \@url }%
\providecommand \@url [1]{\endgroup\@href {#1}{\urlprefix }}%
\providecommand \urlprefix  [0]{URL }%
\providecommand \Eprint [0]{\href }%
\providecommand \doibase [0]{http://dx.doi.org/}%
\providecommand \selectlanguage [0]{\@gobble}%
\providecommand \bibinfo  [0]{\@secondoftwo}%
\providecommand \bibfield  [0]{\@secondoftwo}%
\providecommand \translation [1]{[#1]}%
\providecommand \BibitemOpen [0]{}%
\providecommand \bibitemStop [0]{}%
\providecommand \bibitemNoStop [0]{.\EOS\space}%
\providecommand \EOS [0]{\spacefactor3000\relax}%
\providecommand \BibitemShut  [1]{\csname bibitem#1\endcsname}%
\let\auto@bib@innerbib\@empty
\bibitem [{\citenamefont {Jacox}(2003)}]{Jacox2003}%
  \BibitemOpen
  \bibfield  {author} {\bibinfo {author} {\bibfnamefont {M.~E.}\ \bibnamefont
  {Jacox}},\ }\href@noop {} {\bibfield  {journal} {\bibinfo  {journal} {J.
  Phys. Chem. Ref. Data}\ }\textbf {\bibinfo {volume} {{\bf 32}}},\ \bibinfo
  {pages} {1} (\bibinfo {year} {2003})}\BibitemShut {NoStop}%
\bibitem [{\citenamefont {Sassara}\ \emph {et~al.}(1996)\citenamefont
  {Sassara}, \citenamefont {Zerza},\ and\ \citenamefont
  {Chergui}}]{Sassara1996}%
  \BibitemOpen
  \bibfield  {author} {\bibinfo {author} {\bibfnamefont {A.}~\bibnamefont
  {Sassara}}, \bibinfo {author} {\bibfnamefont {G.}~\bibnamefont {Zerza}}, \
  and\ \bibinfo {author} {\bibfnamefont {M.}~\bibnamefont {Chergui}},\
  }\href@noop {} {\bibfield  {journal} {\bibinfo  {journal} {J. Phys. B}\
  }\textbf {\bibinfo {volume} {{\bf 29}}},\ \bibinfo {pages} {4997} (\bibinfo
  {year} {1996})}\BibitemShut {NoStop}%
\bibitem [{\citenamefont {Salama}\ and\ \citenamefont
  {Allamandola}(1991)}]{Salama1991}%
  \BibitemOpen
  \bibfield  {author} {\bibinfo {author} {\bibfnamefont {F.}~\bibnamefont
  {Salama}}\ and\ \bibinfo {author} {\bibfnamefont {L.}~\bibnamefont
  {Allamandola}},\ }\href@noop {} {\bibfield  {journal} {\bibinfo  {journal}
  {J. Chem. Phys.}\ }\textbf {\bibinfo {volume} {{\bf 94}}},\ \bibinfo {pages}
  {6964} (\bibinfo {year} {1991})}\BibitemShut {NoStop}%
\bibitem [{\citenamefont {Salama}\ \emph {et~al.}(1994)\citenamefont {Salama},
  \citenamefont {Joblin},\ and\ \citenamefont {Allamandola}}]{salama1994}%
  \BibitemOpen
  \bibfield  {author} {\bibinfo {author} {\bibfnamefont {F.}~\bibnamefont
  {Salama}}, \bibinfo {author} {\bibfnamefont {C.}~\bibnamefont {Joblin}}, \
  and\ \bibinfo {author} {\bibfnamefont {L.~J.}\ \bibnamefont {Allamandola}},\
  }\href {\doibase 10.1063/1.467905} {\bibfield  {journal} {\bibinfo  {journal}
  {The Journal of Chemical Physics}\ }\textbf {\bibinfo {volume} {101}},\
  \bibinfo {pages} {10252} (\bibinfo {year} {1994})}\BibitemShut {NoStop}%
\bibitem [{\citenamefont {Halasinski}\ \emph {et~al.}(2000)\citenamefont
  {Halasinski}, \citenamefont {Hudgins}, \citenamefont {Salama}, \citenamefont
  {Allamandola},\ and\ \citenamefont {Bally}}]{Halasinski2000}%
  \BibitemOpen
  \bibfield  {author} {\bibinfo {author} {\bibfnamefont {T.}~\bibnamefont
  {Halasinski}}, \bibinfo {author} {\bibfnamefont {D.}~\bibnamefont {Hudgins}},
  \bibinfo {author} {\bibfnamefont {F.}~\bibnamefont {Salama}}, \bibinfo
  {author} {\bibfnamefont {L.}~\bibnamefont {Allamandola}}, \ and\ \bibinfo
  {author} {\bibfnamefont {T.}~\bibnamefont {Bally}},\ }\href@noop {}
  {\bibfield  {journal} {\bibinfo  {journal} {J. Phys. Chem. A}\ }\textbf
  {\bibinfo {volume} {{\bf 104}}},\ \bibinfo {pages} {7484} (\bibinfo {year}
  {2000})}\BibitemShut {NoStop}%
\bibitem [{\citenamefont {Crépin}\ and\ \citenamefont
  {Tramer}(1990)}]{crepin1990}%
  \BibitemOpen
  \bibfield  {author} {\bibinfo {author} {\bibfnamefont {C.}~\bibnamefont
  {Crépin}}\ and\ \bibinfo {author} {\bibfnamefont {A.}~\bibnamefont
  {Tramer}},\ }\href@noop {} {\bibfield  {journal} {\bibinfo  {journal} {Chem.
  Phys. Lett.}\ }\textbf {\bibinfo {volume} {{\bf 170}}},\ \bibinfo {pages}
  {446} (\bibinfo {year} {1990})}\BibitemShut {NoStop}%
\bibitem [{\citenamefont {Chillier}\ \emph {et~al.}(2001)\citenamefont
  {Chillier}, \citenamefont {Boulet}, \citenamefont {Chermette}, \citenamefont
  {Salama},\ and\ \citenamefont {Weber}}]{Chillier2001}%
  \BibitemOpen
  \bibfield  {author} {\bibinfo {author} {\bibfnamefont {X.}~\bibnamefont
  {Chillier}}, \bibinfo {author} {\bibfnamefont {P.}~\bibnamefont {Boulet}},
  \bibinfo {author} {\bibfnamefont {H.}~\bibnamefont {Chermette}}, \bibinfo
  {author} {\bibfnamefont {F.}~\bibnamefont {Salama}}, \ and\ \bibinfo {author}
  {\bibfnamefont {J.}~\bibnamefont {Weber}},\ }\href@noop {} {\bibfield
  {journal} {\bibinfo  {journal} {J. Chem. Phys.}\ }\textbf {\bibinfo {volume}
  {\bf 115}},\ \bibinfo {pages} {1769} (\bibinfo {year} {2001})}\BibitemShut
  {NoStop}%
\bibitem [{\citenamefont {Joblin}\ \emph {et~al.}(1995)\citenamefont {Joblin},
  \citenamefont {Salama},\ and\ \citenamefont {Allamandola}}]{Joblin1995}%
  \BibitemOpen
  \bibfield  {author} {\bibinfo {author} {\bibfnamefont {C.}~\bibnamefont
  {Joblin}}, \bibinfo {author} {\bibfnamefont {F.}~\bibnamefont {Salama}}, \
  and\ \bibinfo {author} {\bibfnamefont {L.}~\bibnamefont {Allamandola}},\
  }\href@noop {} {\bibfield  {journal} {\bibinfo  {journal} {J. Chem. Phys.}\
  }\textbf {\bibinfo {volume} {{\bf 102}}},\ \bibinfo {pages} {9743} (\bibinfo
  {year} {1995})}\BibitemShut {NoStop}%
\bibitem [{\citenamefont {Joblin}\ \emph {et~al.}(1999)\citenamefont {Joblin},
  \citenamefont {Salama},\ and\ \citenamefont {Allamandola}}]{Joblin1999}%
  \BibitemOpen
  \bibfield  {author} {\bibinfo {author} {\bibfnamefont {C.}~\bibnamefont
  {Joblin}}, \bibinfo {author} {\bibfnamefont {F.}~\bibnamefont {Salama}}, \
  and\ \bibinfo {author} {\bibfnamefont {L.}~\bibnamefont {Allamandola}},\
  }\href@noop {} {\bibfield  {journal} {\bibinfo  {journal} {J. Chem. Phys.}\
  }\textbf {\bibinfo {volume} {{\bf 110}}},\ \bibinfo {pages} {7287} (\bibinfo
  {year} {1999})}\BibitemShut {NoStop}%
\bibitem [{\citenamefont {Almond}\ and\ \citenamefont
  {Orrin}(1991)}]{Almond1991}%
  \BibitemOpen
  \bibfield  {author} {\bibinfo {author} {\bibfnamefont {M.~J.}\ \bibnamefont
  {Almond}}\ and\ \bibinfo {author} {\bibfnamefont {R.~H.}\ \bibnamefont
  {Orrin}},\ }\href@noop {} {\bibfield  {journal} {\bibinfo  {journal} {Annu.
  Rep. Prog. Chem., Sect. C: Phys. Chem.}\ }\textbf {\bibinfo {volume} {88}},\
  \bibinfo {pages} {3} (\bibinfo {year} {1991})}\BibitemShut {NoStop}%
\bibitem [{\citenamefont {Bondybey}\ \emph {et~al.}(1996)\citenamefont
  {Bondybey}, \citenamefont {Smith},\ and\ \citenamefont
  {Agreiter}}]{Bondybey1996}%
  \BibitemOpen
  \bibfield  {author} {\bibinfo {author} {\bibfnamefont {V.}~\bibnamefont
  {Bondybey}}, \bibinfo {author} {\bibfnamefont {A.}~\bibnamefont {Smith}}, \
  and\ \bibinfo {author} {\bibfnamefont {J.}~\bibnamefont {Agreiter}},\
  }\href@noop {} {\bibfield  {journal} {\bibinfo  {journal} {Chem. Rev.}\
  }\textbf {\bibinfo {volume} {\bf 96}},\ \bibinfo {pages} {2113} (\bibinfo
  {year} {1996})}\BibitemShut {NoStop}%
\bibitem [{\citenamefont {Khriachtchev}(2011)}]{Khriachtchev2011}%
  \BibitemOpen
  \bibinfo {editor} {\bibfnamefont {L.}~\bibnamefont {Khriachtchev}},\ ed.,\
  \href@noop {} {\emph {\bibinfo {title} {Physics and Chemistry at Low
  Temperatures}}}\ (\bibinfo  {publisher} {Pan Stanford Publishing},\ \bibinfo
  {year} {2011})\BibitemShut {NoStop}%
\bibitem [{\citenamefont {Dvorak}\ \emph {et~al.}(2012)\citenamefont {Dvorak},
  \citenamefont {M\"uller}, \citenamefont {Knoblauch}, \citenamefont
  {B\"unermann}, \citenamefont {Rydlo}, \citenamefont {Minniberger},
  \citenamefont {Harbich},\ and\ \citenamefont {Stienkemeier}}]{Dvorak_P1}%
  \BibitemOpen
  \bibfield  {author} {\bibinfo {author} {\bibfnamefont {M.}~\bibnamefont
  {Dvorak}}, \bibinfo {author} {\bibfnamefont {M.}~\bibnamefont {M\"uller}},
  \bibinfo {author} {\bibfnamefont {T.}~\bibnamefont {Knoblauch}}, \bibinfo
  {author} {\bibfnamefont {O.}~\bibnamefont {B\"unermann}}, \bibinfo {author}
  {\bibfnamefont {A.}~\bibnamefont {Rydlo}}, \bibinfo {author} {\bibfnamefont
  {S.}~\bibnamefont {Minniberger}}, \bibinfo {author} {\bibfnamefont
  {W.}~\bibnamefont {Harbich}}, \ and\ \bibinfo {author} {\bibfnamefont
  {F.}~\bibnamefont {Stienkemeier}},\ }\href@noop {} {\bibfield  {journal}
  {\bibinfo  {journal} {to be published}\ } (\bibinfo {year}
  {2012})}\BibitemShut {NoStop}%
\bibitem [{\citenamefont {Lallement}\ \emph {et~al.}(1992)\citenamefont
  {Lallement}, \citenamefont {Cuvellier}, \citenamefont {Mestdagh},
  \citenamefont {Meynadier}, \citenamefont {de~Pujo}, \citenamefont
  {Sublemontier}, \citenamefont {Visticot}, \citenamefont {Berlande},\ and\
  \citenamefont {Biquard}}]{Lallement1992}%
  \BibitemOpen
  \bibfield  {author} {\bibinfo {author} {\bibfnamefont {A.}~\bibnamefont
  {Lallement}}, \bibinfo {author} {\bibfnamefont {J.}~\bibnamefont
  {Cuvellier}}, \bibinfo {author} {\bibfnamefont {J.}~\bibnamefont {Mestdagh}},
  \bibinfo {author} {\bibfnamefont {P.}~\bibnamefont {Meynadier}}, \bibinfo
  {author} {\bibfnamefont {P.}~\bibnamefont {de~Pujo}}, \bibinfo {author}
  {\bibfnamefont {O.}~\bibnamefont {Sublemontier}}, \bibinfo {author}
  {\bibfnamefont {J.}~\bibnamefont {Visticot}}, \bibinfo {author}
  {\bibfnamefont {J.}~\bibnamefont {Berlande}}, \ and\ \bibinfo {author}
  {\bibfnamefont {X.}~\bibnamefont {Biquard}},\ }\href@noop {} {\bibfield
  {journal} {\bibinfo  {journal} {Chem. Phys. Lett.}\ }\textbf {\bibinfo
  {volume} {{\bf 189}}},\ \bibinfo {pages} {182} (\bibinfo {year}
  {1992})}\BibitemShut {NoStop}%
\bibitem [{\citenamefont {Briant}\ \emph {et~al.}(2000)\citenamefont {Briant},
  \citenamefont {Gaveau}, \citenamefont {Mestdagh},\ and\ \citenamefont
  {Visticot}}]{Briant2000}%
  \BibitemOpen
  \bibfield  {author} {\bibinfo {author} {\bibfnamefont {M.}~\bibnamefont
  {Briant}}, \bibinfo {author} {\bibfnamefont {M.~A.}\ \bibnamefont {Gaveau}},
  \bibinfo {author} {\bibfnamefont {J.~M.}\ \bibnamefont {Mestdagh}}, \ and\
  \bibinfo {author} {\bibfnamefont {J.~P.}\ \bibnamefont {Visticot}},\ } {\bibfield  {journal} {\bibinfo  {journal} {J. Chem. Phys.}\ }\textbf
{\bibinfo {volume} {112}},\ \bibinfo
  {pages} {1744} (\bibinfo {year} {2000})}\BibitemShut {NoStop}%
\bibitem [{\citenamefont {Briant}\ \emph {et~al.}(2010)\citenamefont {Briant},
  \citenamefont {Gaveau},\ and\ \citenamefont {Mestdagh}}]{Briant2010}%
  \BibitemOpen
  \bibfield  {author} {\bibinfo {author} {\bibfnamefont {M.}~\bibnamefont
  {Briant}}, \bibinfo {author} {\bibfnamefont {M.-A.}\ \bibnamefont {Gaveau}},
  \ and\ \bibinfo {author} {\bibfnamefont {J.-M.}\ \bibnamefont {Mestdagh}},\
  } {\bibfield  {journal} {\bibinfo
  {journal} {J. Chem. Phys.}\ }\textbf {\bibinfo {volume}
  {133}},\ \bibinfo {eid} {034306} (\bibinfo {year} {2010})}\BibitemShut
  {NoStop}%
\bibitem [{\citenamefont {Schreiber}(2000)}]{Schreiber2000}%
  \BibitemOpen
  \bibfield  {author} {\bibinfo {author} {\bibfnamefont {F.}~\bibnamefont
  {Schreiber}},\ }\href@noop {} {\bibfield  {journal} {\bibinfo  {journal}
  {Progress in Surface Science}\ }\textbf {\bibinfo {volume} {{\bf 65}}},\
  \bibinfo {pages} {151} (\bibinfo {year} {2000})}\BibitemShut {NoStop}%
\bibitem [{\citenamefont {Wang}\ and\ \citenamefont {Hersam}(2009)}]{Qing2009}%
  \BibitemOpen
  \bibfield  {author} {\bibinfo {author} {\bibfnamefont {Q.~H.}\ \bibnamefont
  {Wang}}\ and\ \bibinfo {author} {\bibfnamefont {M.}~\bibnamefont {Hersam}},\
  }\href@noop {} {\bibfield  {journal} {\bibinfo  {journal} {Nature Chemistry}\
  }\textbf {\bibinfo {volume} {{\bf 1}}},\ \bibinfo {pages} {206} (\bibinfo
  {year} {2009})}\BibitemShut {NoStop}%
\bibitem [{\citenamefont {Wewer}\ and\ \citenamefont
  {Stienkemeier}(2003)}]{Wewer2003}%
  \BibitemOpen
  \bibfield  {author} {\bibinfo {author} {\bibfnamefont {M.}~\bibnamefont
  {Wewer}}\ and\ \bibinfo {author} {\bibfnamefont {F.}~\bibnamefont
  {Stienkemeier}},\ }\href@noop {} {\bibfield  {journal} {\bibinfo  {journal}
  {Phys. Rev. B}\ }\textbf {\bibinfo {volume} {{\bf 67}}},\ \bibinfo {pages}
  {125201} (\bibinfo {year} {2003})}\BibitemShut {NoStop}%
\bibitem [{\citenamefont {Wewer}\ and\ \citenamefont
  {Stienkemeier}(2004)}]{Wewer2004}%
  \BibitemOpen
  \bibfield  {author} {\bibinfo {author} {\bibfnamefont {M.}~\bibnamefont
  {Wewer}}\ and\ \bibinfo {author} {\bibfnamefont {F.}~\bibnamefont
  {Stienkemeier}},\ }\href@noop {} {\bibfield  {journal} {\bibinfo  {journal}
  {J. Chem. Phys.}\ }\textbf {\bibinfo {volume} {{\bf 120}}},\ \bibinfo {pages}
  {1239} (\bibinfo {year} {2004})}\BibitemShut {NoStop}%
\bibitem [{\citenamefont {Wewer}\ and\ \citenamefont
  {Stienkemeier}(2005)}]{Wewer2005}%
  \BibitemOpen
  \bibfield  {author} {\bibinfo {author} {\bibfnamefont {M.}~\bibnamefont
  {Wewer}}\ and\ \bibinfo {author} {\bibfnamefont {F.}~\bibnamefont
  {Stienkemeier}},\ }\href@noop {} {\bibfield  {journal} {\bibinfo  {journal}
  {Phys. Chem. Chem. Phys.}\ }\textbf {\bibinfo {volume} {{\bf 7}}},\ \bibinfo
  {pages} {1171} (\bibinfo {year} {2005})}\BibitemShut {NoStop}%
\bibitem [{\citenamefont {Roden}\ \emph {et~al.}(2011)\citenamefont {Roden},
  \citenamefont {Eisfeld}, \citenamefont {Dvorak}, \citenamefont {Bünermann},\
  and\ \citenamefont {Stienkemeier}}]{Roden2010}%
  \BibitemOpen
  \bibfield  {author} {\bibinfo {author} {\bibfnamefont {J.}~\bibnamefont
  {Roden}}, \bibinfo {author} {\bibfnamefont {A.}~\bibnamefont {Eisfeld}},
  \bibinfo {author} {\bibfnamefont {M.}~\bibnamefont {Dvorak}}, \bibinfo
  {author} {\bibfnamefont {O.}~\bibnamefont {Bünermann}}, \ and\ \bibinfo
  {author} {\bibfnamefont {F.}~\bibnamefont {Stienkemeier}},\ }\href@noop {}
  {\bibfield  {journal} {\bibinfo  {journal} {J. Chem. Phys}\ }\textbf
  {\bibinfo {volume} {{\bf 134}}},\ \bibinfo {pages} {054907} (\bibinfo {year}
  {2011})}\BibitemShut {NoStop}%
\bibitem [{\citenamefont {Grebenev}\ \emph {et~al.}(1998)\citenamefont
  {Grebenev}, \citenamefont {Toennies},\ and\ \citenamefont
  {Vilesov}}]{Grebenev1998}%
  \BibitemOpen
  \bibfield  {author} {\bibinfo {author} {\bibfnamefont {S.}~\bibnamefont
  {Grebenev}}, \bibinfo {author} {\bibfnamefont {J.}~\bibnamefont {Toennies}},
  \ and\ \bibinfo {author} {\bibfnamefont {A.}~\bibnamefont {Vilesov}},\
  }\href@noop {} {\bibfield  {journal} {\bibinfo  {journal} {Science}\ }\textbf
  {\bibinfo {volume} {{\bf 279}}},\ \bibinfo {pages} {2083} (\bibinfo {year}
  {1998})}\BibitemShut {NoStop}%
\bibitem [{\citenamefont {Scholz}\ \emph {et~al.}(2000)\citenamefont {Scholz},
  \citenamefont {Kobitski}, \citenamefont {Kampen}, \citenamefont {Schreiber},
  \citenamefont {Zahn}, \citenamefont {Jungnickel}, \citenamefont {Elstner},
  \citenamefont {Sternberg},\ and\ \citenamefont {Frauenheim}}]{scholz_raman}%
  \BibitemOpen
  \bibfield  {author} {\bibinfo {author} {\bibfnamefont {R.}~\bibnamefont
  {Scholz}}, \bibinfo {author} {\bibfnamefont {A.}~\bibnamefont {Kobitski}},
  \bibinfo {author} {\bibfnamefont {T.}~\bibnamefont {Kampen}}, \bibinfo
  {author} {\bibfnamefont {M.}~\bibnamefont {Schreiber}}, \bibinfo {author}
  {\bibfnamefont {D.}~\bibnamefont {Zahn}}, \bibinfo {author} {\bibfnamefont
  {G.}~\bibnamefont {Jungnickel}}, \bibinfo {author} {\bibfnamefont
  {M.}~\bibnamefont {Elstner}}, \bibinfo {author} {\bibfnamefont
  {M.}~\bibnamefont {Sternberg}}, \ and\ \bibinfo {author} {\bibfnamefont
  {T.}~\bibnamefont {Frauenheim}},\ }\href {\doibase 10.1103/PhysRevB.61.13659}
  {\bibfield  {journal} {\bibinfo  {journal} {Phys. Rev. B}\ }\textbf {\bibinfo
  {volume} {{\bf61}}},\ \bibinfo {pages} {13659} (\bibinfo {year}
  {2000})}\BibitemShut {NoStop}%
\bibitem [{\citenamefont {Lehnig}\ and\ \citenamefont
  {Slenczka}(2003)}]{lehnig2003}%
  \BibitemOpen
  \bibfield  {author} {\bibinfo {author} {\bibfnamefont {R.}~\bibnamefont
  {Lehnig}}\ and\ \bibinfo {author} {\bibfnamefont {A.}~\bibnamefont
  {Slenczka}},\ }\href@noop {} {\bibfield  {journal} {\bibinfo  {journal} {J.
  Chem. Phys.}\ }\textbf {\bibinfo {volume} {{\bf 118}}},\ \bibinfo {pages}
  {8256} (\bibinfo {year} {2003})}\BibitemShut {NoStop}%
\bibitem [{\citenamefont {Bünermann}(2006)}]{Buenermann2006}%
  \BibitemOpen
  \bibfield  {author} {\bibinfo {author} {\bibfnamefont {O.}~\bibnamefont
  {Bünermann}},\ }\emph {\bibinfo {title} {Spektroskopie von Alkali- und
  Erdalkaliatomen, -molekülen, Alkaliclustern und Komplexen organischer
  Moleküle auf Heliumnanotröpfchen}},\ \href@noop {} {Ph.D. thesis},\ \bibinfo
  {school} {Universität Bielefeld} (\bibinfo {year} {2006})\BibitemShut
  {NoStop}%
\bibitem [{\citenamefont {Imasaka}\ \emph {et~al.}(1984)\citenamefont
  {Imasaka}, \citenamefont {Fukuoka}, \citenamefont {Hayashi},\ and\
  \citenamefont {Ishibashi}}]{Imasaka1984}%
  \BibitemOpen
  \bibfield  {author} {\bibinfo {author} {\bibfnamefont {T.}~\bibnamefont
  {Imasaka}}, \bibinfo {author} {\bibfnamefont {H.}~\bibnamefont {Fukuoka}},
  \bibinfo {author} {\bibfnamefont {T.}~\bibnamefont {Hayashi}}, \ and\
  \bibinfo {author} {\bibfnamefont {N.}~\bibnamefont {Ishibashi}},\ }\href@noop
  {} {\bibfield  {journal} {\bibinfo  {journal} {Analytica Chimica Acta}\
  }\textbf {\bibinfo {volume} {{\bf 156}}},\ \bibinfo {pages} {111} (\bibinfo
  {year} {1984})}\BibitemShut {NoStop}%
\bibitem [{\citenamefont {Callegari}\ \emph {et~al.}(1998)\citenamefont
  {Callegari}, \citenamefont {Higgins}, \citenamefont {Stienkemeier},\ and\
  \citenamefont {Scoles}}]{Callegari1998}%
  \BibitemOpen
  \bibfield  {author} {\bibinfo {author} {\bibfnamefont {C.}~\bibnamefont
  {Callegari}}, \bibinfo {author} {\bibfnamefont {J.}~\bibnamefont {Higgins}},
  \bibinfo {author} {\bibfnamefont {F.}~\bibnamefont {Stienkemeier}}, \ and\
  \bibinfo {author} {\bibfnamefont {G.}~\bibnamefont {Scoles}},\ }\href@noop {}
  {\bibfield  {journal} {\bibinfo  {journal} {J. Phys. Chem. A}\ }\textbf
  {\bibinfo {volume} {102}},\ \bibinfo {pages} {95} (\bibinfo {year}
  {1998})}\BibitemShut {NoStop}%
\bibitem [{\citenamefont {Biktchantaev}\ \emph {et~al.}(2002)\citenamefont
  {Biktchantaev}, \citenamefont {Samartsev},\ and\ \citenamefont
  {Sepiol}}]{Biktchantaev2002}%
  \BibitemOpen
  \bibfield  {author} {\bibinfo {author} {\bibfnamefont {I.}~\bibnamefont
  {Biktchantaev}}, \bibinfo {author} {\bibfnamefont {V.}~\bibnamefont
  {Samartsev}}, \ and\ \bibinfo {author} {\bibfnamefont {J.}~\bibnamefont
  {Sepiol}},\ }\href@noop {} {\bibfield  {journal} {\bibinfo  {journal}
  {Journal of Luminescence}\ }\textbf {\bibinfo {volume} {98}},\ \bibinfo
  {pages} {265} (\bibinfo {year} {2002})}\BibitemShut {NoStop}%
\bibitem [{\citenamefont {Klein}\ and\ \citenamefont
  {Venables}(1976)}]{Klein1976}%
  \BibitemOpen
  \bibfield  {author} {\bibinfo {author} {\bibfnamefont {M.}~\bibnamefont
  {Klein}}\ and\ \bibinfo {author} {\bibfnamefont {J.}~\bibnamefont
  {Venables}},\ }\href@noop {} {\emph {\bibinfo {title} {Rare gas solids}}}\
  (\bibinfo  {publisher} {Academic Press},\ \bibinfo {year} {1976})\BibitemShut
  {NoStop}%
\end{thebibliography}

%


\end{document}